\documentclass[aps,twocolumn,nopacs,preprintnumbers]{revtex4}
\usepackage{graphicx}
\usepackage{bm}
\usepackage{color}
\usepackage{amsmath}
\usepackage{float}
\usepackage{blindtext}
\usepackage{epsfig}
\usepackage{epstopdf}
\usepackage{multirow}
\usepackage{comment}
\usepackage{latexsym}
\usepackage{mathrsfs}
\usepackage{natbib}
\usepackage[colorlinks=true,
            linkcolor = blue,
            urlcolor  = blue,
            citecolor = blue,
            anchorcolor = blue]{hyperref}
\usepackage{url}
\usepackage{graphicx}
\usepackage[normalem]{ulem}
\usepackage{xcolor}

\newcommand\redsout{\bgroup\markoverwith{\textcolor{red}{\rule[0.5ex]{2pt}{0.4pt}}}\ULon}

\newcommand{\beq}{\begin{equation}}
\newcommand{\eeq}{\end{equation}}
\newcommand{\bea}{\begin{eqnarray}}
\newcommand{\eea}{\end{eqnarray}}

\usepackage{calligra}
\DeclareMathAlphabet{\mathcalligra}{T1}{calligra}{e}{f}

\begin{document}
\title{MoB$_2$ under Pressure: Superconducting Mo Enhanced by Boron}
\author{Yundi Quan}
\affiliation{Department of Physics, University of Florida, Gainesville, Florida 32611, USA}
\affiliation{Department of Materials Science and  Engineering, University of Florida, Gainesville, Florida 32611, USA}
\affiliation{Quantum Theory Project, University of Florida, Gainesville, Florida 32611, USA}
\author{Kwan-Woo Lee}
\affiliation{Division of Display and Semiconductor Physics, Korea University, Sejong 30019, Korea}
\affiliation{Department of Applied Physics, Graduate School, Korea University, Sejong 30019, Korea}
\author{Warren E. Pickett}
\affiliation{Department of Physics and Astronomy, University of California, Davis, California 95616, USA}

\date{\today}

\begin{abstract}
The discovery of the first high critical temperature (T$_c$) transition metal diboride superconductor, MgB$_2$ structure $\alpha$-MoB$_2$ under pressure with T$_c$ up to 32 K at 100 GPa, provides new input into some unexplained aspects of electron-phonon coupling in intermetallic compounds.  We establish that MoB$_2$ is a phonon-mediated superconductor but has little in common with MgB$_2$ (T$_c$=40 K at zero pressure). MoB$_2$ is a strongly metallic, three dimensional, multi-Fermi surface material, becoming of additional interest because it displays a frequency separation of Mo and B vibrations that mirrors that of metal superhydrides with T$_c$ approaching room temperature. This separation, which is unusual in intermetallic compounds, allows an analysis separately for Mo and B providing, amongst the other parameters essential for understanding phonon coupling, the matrix elements for scattering by the individual atoms. Strong coupling ($\lambda_{\rm Mo}$=1.48) is provided by Mo (total $\lambda=$1.67). A factor of 15 weaker coupling to each B atom is compensated by that coupling being to mean high frequency modes around 85-90 meV (maximum of 140 meV), versus 18-20 meV for Mo. As a result, B enhances T$_c$ by  
43\% over the Mo-only value, to 33 K, corresponding to the experimental value.  These results provide a guideline for designing higher T$_c$ materials from a cooperation of strong coupling from heavy atoms with weakly coupled light atoms.  The new high T$_c$ paradigm discovered here highlights the need for studying and engineering larger ionic scattering matrix elements. 
\end{abstract}
\maketitle

\section{Background}
The discovery two decades ago\cite{Nagamatsu2001} of superconductivity in MgB$_2$ at 40 K reinvigorated interest in electron-phonon (EP) coupled superconductors. The maximum at that time, after 20 years of very active research in the 1960s and 1970s, was 23 K, and there were suggestions that 30 K would be the limit for the EP mechanism. Perhaps more remarkable was that the previous highest T$_c$ materials had been high symmetry (essentially cubic) transition metal dominated compounds, whereas MgB$_2$ is an $s$-$p$ metal with two dimensional (2D) Fermi surfaces that provide the high T$_c$ carriers \cite{An2001,Kortus2001,Bohnen2001,Kong2001,Liu2001,Choi2002,Yildirim2002,Choi2002b}. MgB$_2$ broke nearly all of the ``Matthias rules'' of supposed HTS. The breakthrough in cuprates (largely a non-EP mechanism) broke the final Matthias rule by requiring oxygen to be a crucial actor in the electronic and magnetic structure.

The remarkable example of MgB$_2$ prompted exploration of other diborides, especially transition metal (TM) diborides which provided the seemingly important elements: a covalent boron sublattice together with TM $d$ bands that can provide metallic-covalent TM-boron bonding and thereby additional coupling. The theoretical (and experimental) results were disappointing:\cite{Rosner2001} disrupting the simple electron donation (to the B honeycomb sublattice) did not lead to increased T$_c$.

The recent discovery \cite{MoB2} of T$_c$ up to 32 K in non-magnetic MoB$_2$, albeit under pressure, should provide avenues to pursue to obtain higher T$_c$ materials, if its lessons can be identified. Useful exploration requires new understanding: what is it in MoB$_2$ that makes it so different from other diborides, and what causes the rapid increase in T$_c$ under pressure, when the majority of superconductors show a decrease in T$_c$ with pressure. MoB$_2$ itself is not superconducting, but is stabilized by 4\% of Zr for Mo that produces stoichiometric materials, with T$_c \approx 6$ K \cite{Muzzy2002}.  MoB$_2$, and several other transition metal borides, are difficult metallurgically \cite{Muzzy2002,Tang2020,Ren2008}, and good prospects in this class have been difficult to find \cite{Rosner2001}.
MoB$_2$ thus provides a new, yet to be understood, paradigm in metal-metalloid superconductivity. This discovery also extends the realization that high pressure can produce high T$_c$ in intermetallic materials, such as up to 19 K above 150 GPa in a Nb-Ti alloy \cite{Guo2019}.

Metal hydrides at high pressure, viz. H$_3$S at 200 K and a few others reported with T$_c$ up to 280 K, introduced their own paradigm \cite{Duan2014,Drozdov2015,Einaga2016,Duan2017,ME2018a,MS2019,ME2018b,Ma2014,Pickett2021}. As anticipated by Ashcroft \cite{Ashcroft}, the hard H vibrational modes provide sufficient coupling to lead to the ultrahigh T$_c$ values approaching room temperature \cite{Duan2014,Ma2014,Papa2015,errea2015,bernstein2015,akashi2015,jose2016,Quan2016}. Calculations based on density functional theory (DFT) for electrons and phonons, and Eliashberg theory for their coupling, accounted very well for the extreme values of $T_c$, but with limited understanding because both metal and H atoms are entangled in the basic material properties (see Sec. II). Initial impressions of the hydrides were that since the metal atom vibrations provide 10-20\% of the EP coupling strength $\lambda$, they contributed similarly to $T_c$. 

More detailed analysis \cite{Quan2019} enabled by the large mass difference of metal and H atoms, whose formalism is presented in Sec. II, shows that the metal modes in hydrides are useless for increasing $T_c$, because their frequencies are low compared to the light atom H vibrations and they reduce the mean phonon frequency. In fact the metal contributions impede the understanding of origin and trends in $T_c$. New analysis revealed that the contribution of the metal atom can even be negative (harmful, but by a small amount) \cite{Quan2019}. What we find in this study, by applying the same analysis, is that MoB$_2$ provides a new paradigm: while the Mo modes and B modes are separated in frequency as in the hydrides, the combination of coupling strength and frequency range provide a new example of how specific pairs of atoms may cooperate to promote high T$_c$, in a compound in which the contributions of each atom type can be disentangled in detail. 

Section II presents the formalism necessary to perform the new (but straightforward) analysis of atom-resolved material characteristics. Section III provides descriptions of the crystal structures that will be studied, followed by the computational methods used in the electronic structure, phonon spectrum and coupling, and analysis. In Sec. IV we provide the main computational results for band structure and Fermi surfaces. The results from density functional theory (DFT) and Eliashberg theory are presented and analyzed in Sec. V, including the atom-specific isotope shifts of T$_c$. Collecting the results to learn why MoB$_2$ is a high temperature superconductor is done in Sec. VI. Section VII provides a summary of our results.   


\section{Formalism}
\subsection{Background}
The critical temperature T$_c$ (and other superconducting properties) in a phonon-coupled superconductor are determined by the frequency $\omega$ resolved Eliashberg spectral function $\alpha^2F(\omega)$ and a Coulomb repulsion parameter $\mu^*$. This latter parameter is around 0.10-0.15 in most superconductors and will not be considered here. $F(\omega)$ is the phonon density of states, and $\alpha^2(\omega)=\alpha^2F(\omega)/F(\omega)$ is a frequency resolved coupling strength. The conventional coupling strength $\lambda$ is the $\omega^{-1}$-weighted moment given by 
\begin{eqnarray}
\lambda=\int \frac{2}{\omega}\alpha^2F(\omega) d\omega.
\end{eqnarray}
Understanding is facilitated by defining the shape function\cite{pba1972,alldyn}
\begin{eqnarray}
g(\omega)=\frac{2}{\lambda}\alpha^2F(\omega)
\end{eqnarray}
whose integral is normalized to unity. Frequency moments $\omega_n$ of the spectrum, independent of the strength $\lambda$ but weighted by coupling strength, are defined by
\begin{eqnarray}
\omega_n=\left\lbrack\int \omega^n g(\omega) d\omega \right \rbrack^\frac{1}{n},
\end{eqnarray}
with the $n\rightarrow0$ limit \cite{alldyn} giving the logarithmic moment $\omega_{log}$. Moments are a non-decreasing function of $n$; for an Einstein spectrum $\delta(\omega-\omega_E)$ all moments are equal to $\omega_E$; the broader the spectrum, the more the moments are separated. 

Except for highly unusual shapes, the first few moments provide a useful characterization of $g(\omega)$. This observation led Allen and Dynes (AD) \cite{alldyn} to fit a generalization of the McMillan equation \cite{McMillan} T$_c^M(\lambda,\omega_2;\mu^*)$ to a more accurate analytic form $T_c^{AD}(\lambda,\omega_{log},\omega_2;\mu^*)$ fitted to 200 experimental or model forms of $\alpha^2F(\omega)$ with known $T_c$. 
We also mention (and use) the observation of Leavens and Carbotte \cite{leavens} that, at least for the strong coupling superconductors of the time, the empirical relation $T_c^{LC}(\lambda,\omega_1)=0.148\lambda \omega_1/2\equiv 0.148 {\cal A}$ reproduced experimental and calculational data at the time; this product involves only the area ${\cal A}$ under $\alpha^2F(\omega)$ \cite{leavenscomment}. It was established by Quan {\it et al.} \cite{Quan2019} that this expression holds well (with the same constant) for metal hydrides of three structural classes, {\it if} the low frequency contribution of the metal atom is neglected. How MoB$_2$ fits into this picture will be discussed in Sec. V.

Much of the early literature addressing how to increase $T_c$ focused on increasing $\lambda$, with a seminal study by Bergmann and Rainer \cite{Rainer} providing a frequency resolved functional derivative $\delta T_c/\delta \alpha^2F(\omega)$ from Eliashberg theory which is non-negative and for standard shapes peaks around $\omega\approx 2\pi T_c$ ($\hbar$=1, $k_B$=1 units). Experimentally, however, one can rarely simply add a $\delta$-function contribution to $\alpha^2F(\omega)$ (as the functional derivative does), instead one makes a change in a material that produces changes of magnitude or even sign in different ranges of frequency. The understanding of how physical changes in $\alpha^2F(\omega)$ affect $T_c$ is a rather subtle undertaking \cite{Quan2019}.

\subsection{The microscopic properties}
From the definition of $\alpha^2F(\omega)$ McMillan obtained for an elemental material \cite{McMillan}, and Hopfield studied \cite{Hopfield}, the decomposition
\begin{eqnarray}
\lambda=\frac{N^{\uparrow}(0) {\cal I}^2}{M\omega_2^2}\equiv\frac{\eta}{\kappa},
\label{eq:lambda}
\end{eqnarray}
where N$^{\uparrow}$(0) is single-spin\cite{alldyn} Fermi level (equal to zero) density of states (DOS), ${\cal I}^2$ is the Fermi surface averaged squared electron-ion matrix element, $M$ is the atomic mass, and $\omega_2$ is the second frequency moment of $g(\omega)$. The scattering strength is given by the change in crystal potential $V(\vec r;\{R\})$ due to the displacement of the atom at $\vec R$
\begin{eqnarray}
{\cal I}^2=\left\langle\left\langle|\langle k|\frac{dV}{d\vec R}|k'\rangle|^2\right\rangle\right\rangle,
\label{eq:matrixelement}
\end{eqnarray}
where the large brackets indicate a double average of $\vec k, \vec k'$ over the Fermi surface, leaving ${\cal I}^2$ independent of $N^{\uparrow}(0)$. It is an atomic property, but material dependent, and the degree and character of its material dependence is one of the fundamental open questions in formulating strategies for increasing $T_c$. Details of the formal theory of electron-phonon coupling, and several aspects of implementation, can be found in the review of Giustino \cite{Giustino2017}.

The final expression in Eq.~(\ref{eq:lambda}) specifies the McMillan-Hopfield quantity $\eta$ and an effective sublattice stiffness $\kappa$ that, like the simple harmonic oscillator, are independent of mass \cite{AllenMitrovic}.  This simple form, with $N^{\uparrow}(0)$ available from calculation and $\omega_2$ from experiment (or in recent years, from DFT), has provided the basis of thinking about how to increase $T_c$. For general compounds, all of these material parameters include contributions from all atom types, which have largely halted quantitative analysis. The Gaspari-Gy\"orffy rigid muffin-tin model for ${\cal I}^2$ has allowed some progress in understanding individual atomic contributions \cite{GG} (see also Sec. VI). 

As mentioned, compounds present challenges in obtaining the relative importance of the various constituent atoms, because each of the quantities in Eq.~(\ref{eq:lambda}) depends on all atomic species in the unit cell. However, pressurized metal hydrides have provided examples that can be handled quantitatively and accurately because the spectra of $F(\omega), \alpha^2F(\omega)$, and $g(\omega)$ separate into disjoint frequency ranges, for the metal atom at low frequency and for H vibrations at high frequency. With excellent accuracy, one obtains expression for $\lambda_j$ for each atom as in Eq.~(\ref{eq:lambda}). Specifically,
\begin{eqnarray}
\lambda_j=\frac{N^{\uparrow}_j(0) {\cal I}_j^2}{M_j \omega_{2,j}^2},~~~ \lambda=\lambda_X + \lambda_H.
\label{eq:atomlambda}
\end{eqnarray}
$N^{\uparrow}(0)$ must be divided into atomic contributions (preferably including the interstitial contribution). From these expressions one obtains the atomic ${\cal I}_j^2$ values, the one central quantity that is in most need for increasing $T_c$ and understanding its variation with respect to changes, viz. pressure. $\lambda$ then is the sum of atomic contributions, and frequency moments can be obtained for each atom. The entire $\alpha^2F(\omega)$ function still provides $\lambda$ and the full frequency moments from all atoms. 

\subsection{Isotope shift}
Not mentioned previously is that the atom-specific isotope shifts of $T_c$ can be obtained trivially. If one atom type, say B, is changed from $^{10}B$ to $^{11}B$, it is only necessary to scale the B frequency spectrum ({\it i.e.} the B moments) by $\sqrt{10/11}$ and reevaluate the frequency moments and then $T_c$. For a general compound, one must replace the atomic mass(es) and recalculate the phonon spectrum and Eliashberg function. (We have neglected here an esoteric mass dependence of $\mu^*$.)

The separation of atomic phonon spectra makes it simple to calculate the isotope shift for each atom, using the Allen-Dynes equation. First, the lattice stiffness $M\omega_2^2$ for each atom does not depend on the isotope. Second, the atomic frequency moments behave in a simple way (since they are independent of other atom masses): $\omega_q \propto M^{-1/2}$ for all modes $q$, so the entire atomic spectrum shifts simply by scaling the frequency.

First, the relationship between the atomic (Mo or B) moments and the net moment is obtained by writing $g=g^{\rm Mo}+g^{\rm B}$ and normalizing each region separately,
\bea
\omega_n^n&=& \int g(\omega)\omega^n d\omega 
   = \int [g^{\rm Mo}(\omega) \omega^n + g^{\rm B}(\omega)\omega^n] d\omega \nonumber \\
   &=&\int_{\rm Mo} \frac{g(\omega)\omega^n d\omega} {\int_{\rm Mo} g(\omega')d\omega'} +
      \int_{\rm B}    \frac{g(\omega)\omega^n d\omega} {\int_{\rm B}    g(\omega')d\omega'}\nonumber \\
   &=& \frac{\lambda^{\rm Mo}}{\lambda}  [\omega_n^{\rm Mo}]^n
    +  \frac{\lambda^{\rm B}} {\lambda}   [\omega_n^{\rm B}]^n.
\eea
The denominator integrals over the Mo region (respectively B region) are needed to provide the correctly normalized values of $\omega_n^{\rm Mo}$ and $\omega_n^{\rm B}$, which involve averages over only their own spectral region. For $\omega_{log}$, it is the logarithm that is averaged:
\bea
{\rm ln}~\omega_{log}&=&\frac{\lambda^{\rm Mo}}{\lambda}{\rm ln}~\omega_{log}^{\rm Mo}
                + \frac{\lambda^{\rm B}} {\lambda}{\rm ln}~\omega_{log}^{\rm B} \nonumber \\
\omega_{log}&=&(\omega_{log}^{\rm Mo})^{\lambda^{\rm Mo}/\lambda}
               (\omega_{log}^ {\rm B})^{\lambda^{\rm B}/\lambda}. 
\eea

With the change in an isotope from mass $M$ to $M'$, $M\omega^2_2$ and electronic properties are unchanged, so $\lambda$ (including each atomic contribution) remains unchanged. The isotope effect arises from how frequency moments appear in T$_c$. We will use the Allen-Dynes equation, which is systematic and simple, but reasonably accurate and widely used. It includes the moment $\omega_{log}$ straightforwardly in the prefactor, and $\omega_2$ in the strong-coupling correction, and their ratio in the shape factor. 

When the isotope of Mo (analogously for B) is decreased by 1 amu from $M$ to $\bar{M}$, the Mo frequency spectrum is stretched by $\sqrt{M/\bar{M}}=\nu$: $\omega\rightarrow \nu\omega.$ The individual moments are a function of the shape in their part of the spectrum; specifically, they are independent of the other atom and of the coupling strengths. With the new mass, the normalized shape function is $G(\omega/\nu)/\nu$. The changes in the moments are immediate:
\bea
{\bar{\omega}_n}=\nu\omega_n:~~~ \bar{\omega}_{log}=\nu\omega_{log}.
\eea

~
\end{comment}

The corresponding $\bar{\omega}_{log}$ and $\bar{\omega}_2$ numbers of Mo must be combined with the B moments using the expressions above, and substituted into the T$_c$ equation. The difference in T$_c$ provides the isotope shift coefficients $\Delta {\rm log}T_c/\Delta {\rm log}M$ for each atom, for the discrete $\Delta M=1$ amu. The results are presented and discussed for $\alpha$-MoB$_2$ in Sec. V E.

\subsection{Earlier application to metal hydrides}
 For binary hydrides $X{\rm H}_n$ ($X$ is a metal atom), the various atom-specific and total quantities were obtained and applied \cite{Quan2019,RHtheory,LaH10theory,eva1,eva2} to great advantage to better understand the origins of their high $T_c$ and provide insight into limits of $T_c$. Three crystal classes encompassing five hydrides were studied. An important point of this formalism is that, given $\alpha^2F(\omega)$, ${\cal I}_j^2$ can be obtained for each atom, and its variation -- for example with pressure -- can be extracted and analyzed. 
 
 This formalism allowed the quantitative importance for $T_c$ of H versus the metal atom. Hydrogen's dominance is, in effect, total. While the low $\omega$ vibrations of the metal provide a significant contribution to $\lambda$ (typically $\sim$15\%), it does not contribute to $T_c$: the increase in $\lambda$ is offset by the decrease in frequency moments, and sometimes the metal contribution to $T_c$ is negative.  Another discovery was that the H value of ${\cal I}^2_{\rm H}$ varied by a factor of five over these hydrides, all high $T_c$ examples. Also, the pressure (volume) variation was not very systematic nor particularly large, although ${\cal I}^2_{\rm H}$ tended to increase with pressure whereas a simple picture involving additional screening due to increased electronic density would suggest the opposite. Various ways to explore correlations between H quantities were presented and discussed. 
 
 Based on early data on H$_3$S and anticipated similar cases with distinct metal and H spectra (similarly, acoustic and optical spectra), Gorkov and Kresin present expressions for T$_c$ in various regimes \cite{Gorkov2018} based on the two corresponding coupling constants $\lambda_{ac}$ and $\lambda_{op}$, and mean frequencies. We do not in this paper try to generalize expressions for T$_c$.  
 A primary aim of this paper is to extend the previous analysis for hydrides \cite{Quan2019} to the metal-metalloid superconductor MoB$_2$, whose $T_c$ is predicted and observed to be in the 30-35 K range at megabar pressures.

\begin{figure}[tbp]
  \includegraphics[width=0.95\columnwidth]{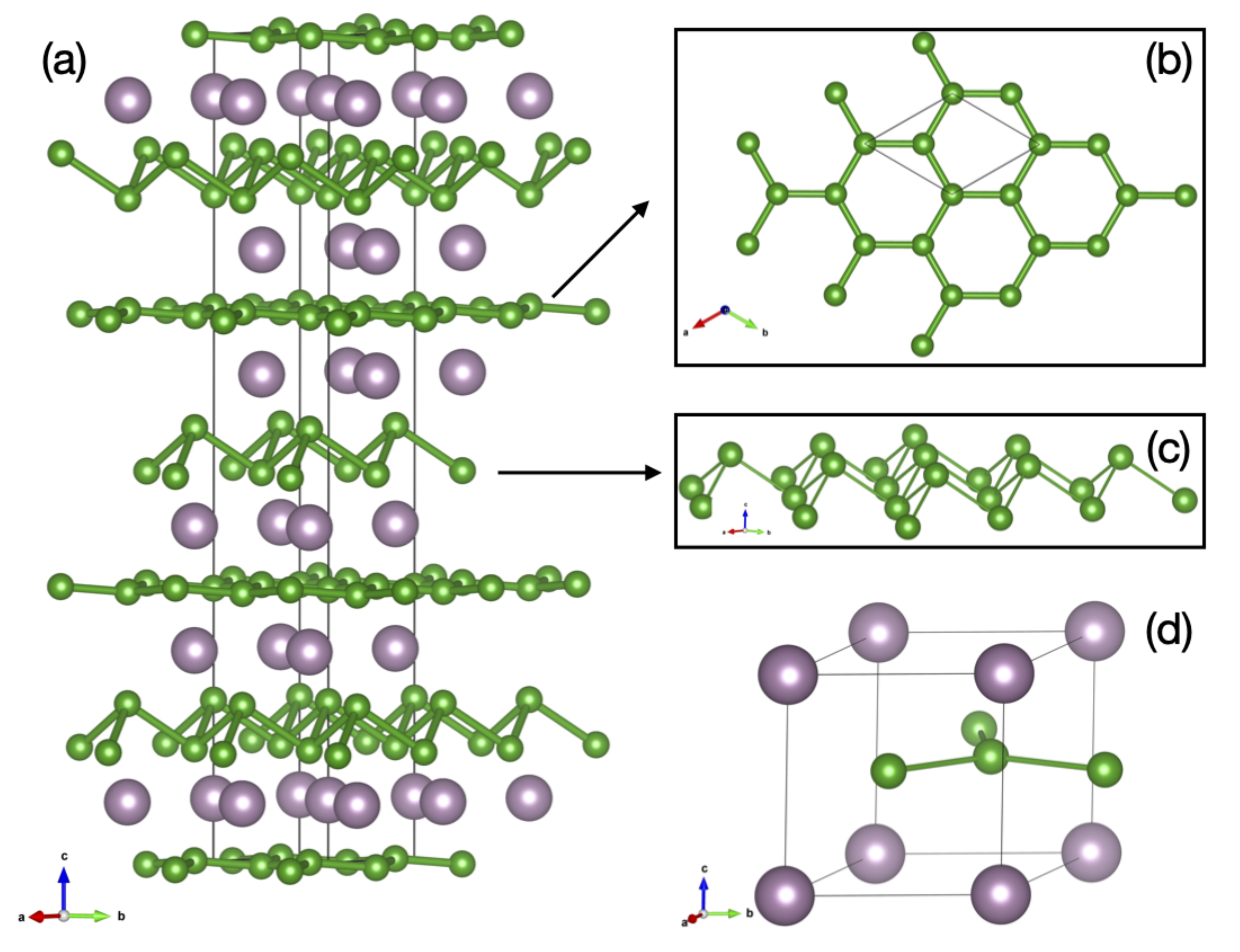}
  \caption{(a) Crystal structure of $\beta$-MoB$_2$, showing the puckering of the B$_6$ layers and the sequential displacement of both layers. (b) The planar B honeycomb layer, with (c) showing another view of the buckled layer. (d) The unit cell of AlB$_2$ structure $\alpha$-MoB$_2$.}
  \label{fig:structure}
\end{figure}

\section{Structure and Methods}
Pressure studies of MoB$_2$ have revealed two polymorphs:\cite{MoB2} $\beta$-MoB$_2$ up to 60 GPa pressure with T$_c$ increasing rapidly from 3 K to 28 K in the 25-60 GPa range, followed by $\alpha$-MoB$_2$ where T$_c$ continues to increase but more slowly, to 32 K at 110 GPa. Both structures include triangular Mo sublattices with honeycomb B$_6$ layers.  The high pressure phase has the well known AlB$_2$-structure of MgB$_2$, symmorphic with inversion. The $\beta$ polymorph, pictured in Fig.~\ref{fig:structure}(a-c), differs in the stacking sequence of graphene structure layers of B$_6$, with symmetry allowing puckering of these planes. Alternate B planes are large and small degrees of puckering.

\begin{table*}[!ht]
  \begin{tabular}{|cc|cc|cccc|cccc|cccccc}
  \hline\hline
$~{\cal P}$ (GPa)~ & ~Phase~ & ~$a$ (\AA )~ & ~$c$ (\AA)~ &~~~& ~$z_{Mo}$~ &  ~$z_{B1}~$ & ~$z_{B2}$~ & ~Mo-B1~ & ~Mo-B2~ & ~B1-B1~ & ~B2-B2~\\
\hline
0.3 & $\beta$ & 3.010 & 21.734 & & 0.0747 & 0.3353 & 0.1729 & 2.349 & 2.134 & 1.740 & 1.759\\
90 & $\beta$ & 2.859 &  19.796 & & 0.0753 & 0.3297 & 0.1944 &  2.177 & 2.076 & 1.657 & 1.983\\
90 & $\alpha$ & 2.884 & [18.102] & &- & - & - &2.247 & 2.247 & 1.665 & 1.665\\
\hline
0.3$^\ast$ & $\beta$ & 3.010 & 21.734 & & 0.0768 & 0.3320 & 0.1817 & 2.390 & 2.281 & 1.739 & 1.857\\
60$^\ast$ & $\beta$ & 2.880 &  20.200 & & 0.0760 & 0.3319 & 0.1827 &  2.242 & 2.159 & 1.664 & 1.785\\
90$^\ast$ & $\beta$ & 2.859 & 19.796  & &0.0756 & 0.3319 & 0.1830 &  2.211 & 2.124 & 1.652 & 1.773\\
\hline\hline
\end{tabular}
\caption{Experimental structure parameters \cite{MoB2} and atomic-bond distances (in \AA) of MoB$_2$ under pressure at room temperature. 
The symbol of $\ast$ at each pressure indicates our relaxed values with the experimental lattice parameters within GGA (LDA values are almost identical to those of GGA). The optimization was carried out until forces were less than 1 meV/\AA~ in {\sc fplo}. In the $\beta$-phase containing two formula units (space group: $R\bar{3}m$, No. 166), all atoms sit at $6c$ (0,0,$z$) sites. In the $\alpha$-phase, the atoms sit at high symmetry AlB$_2$ sites as pictured in Fig. \ref{fig:structure}(d). The Mo-Mo distance is equal to the lattice constant $a$. For the $\alpha$ structure, B1 and B2 are symmetry related, and for comparison, 6$\times$$c$ is presented ($c$=3.017\AA).}
\label{str}
\end{table*}





The experimental structural parameters and atomic distances \cite{MoB2} at 0.3 GPa and 90 GPa are given in Table \ref{str}.
These values show that 60 GPa pressure in the $\beta$ phase distorts the nearly equal B-B distance at low pressure to dimerized values differing by 20\%. In addition, the Mo-B1 distance shortens by almost 8\% while the Mo-B2 distance hardly changes, being 8\% smaller than the Mo-B distance in the $\alpha$-phase at 90 GPa. The structural transition at 60 GPa thus causes a substantial rearrangement in Mo-B and B-B bonding, with its return to the high symmetry AlB$_2$ structure.

The band structure and Fermi surfaces are studied with the full potential, all-electron code {\sc fplo} \cite{fplo} using the semi-local generalized gradient approximation (GGA) \cite{GGA} exchange-correlation functional. This code uses a basis of pre-determined atomic orbital-like functions, with various numerical techniques to handle the full potential and general charge density of the crystal. These calculations use valence and virtual $s$ and $p$ orbitals for B, and valence and virtual $s$ and $p$ orbitals and $4d$ orbitals for Mo.

The pseudopotential code {\sc quantum espresso}  (QE) \cite{QE} is used to carry out linear response electron-phonon calculations. The electron band $k$- and phonon branch $q$-meshes are Monkhorst-Pack type with 18$\times$18$\times$18 and 6$\times$6$\times$6 points, respectively. The optimized norm conserving pseudopotentials \cite{Hamann} were used. 

\begin{figure}[!ht]
  \includegraphics[width=\columnwidth]{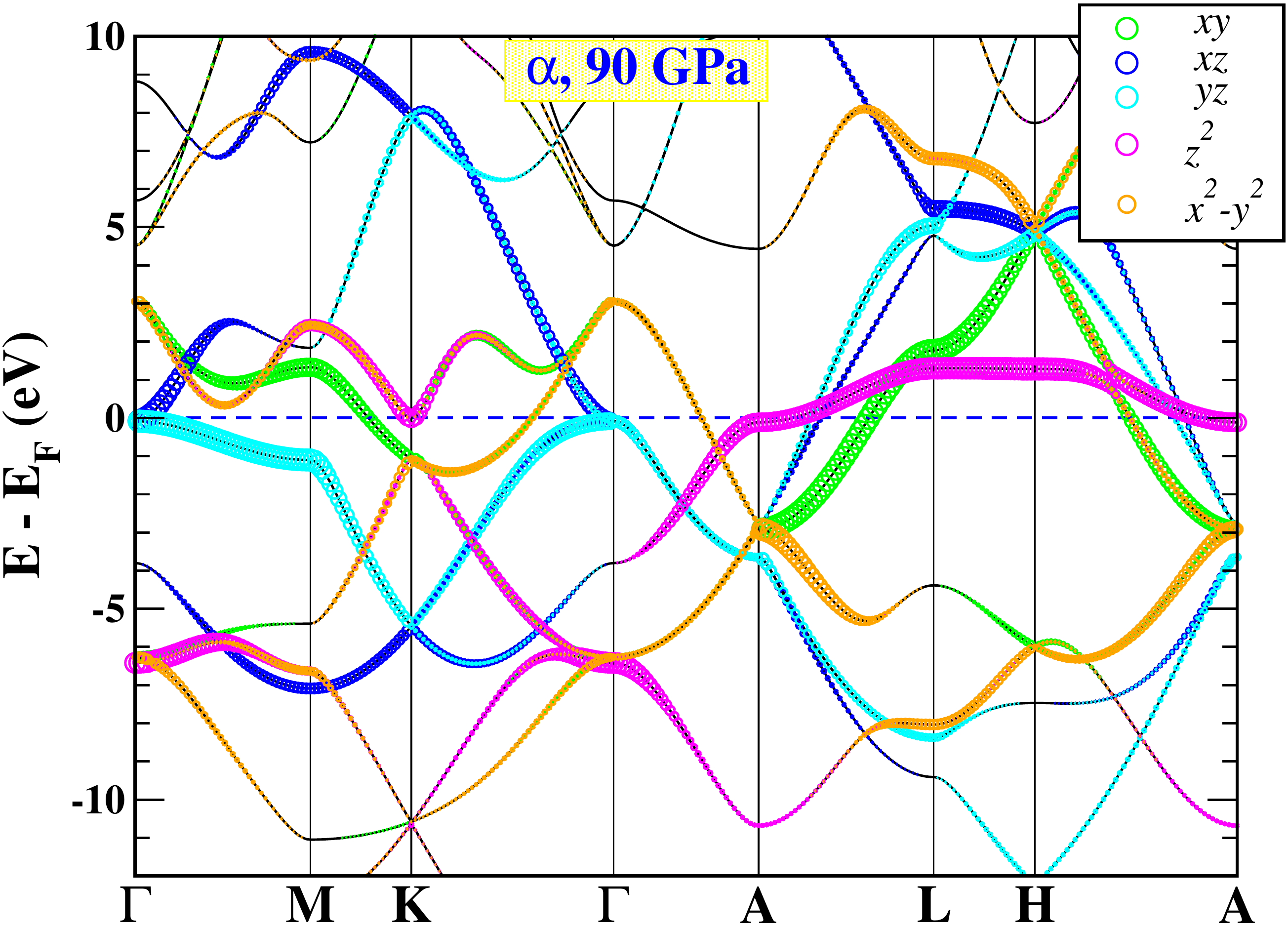}
  \includegraphics[width=\columnwidth]{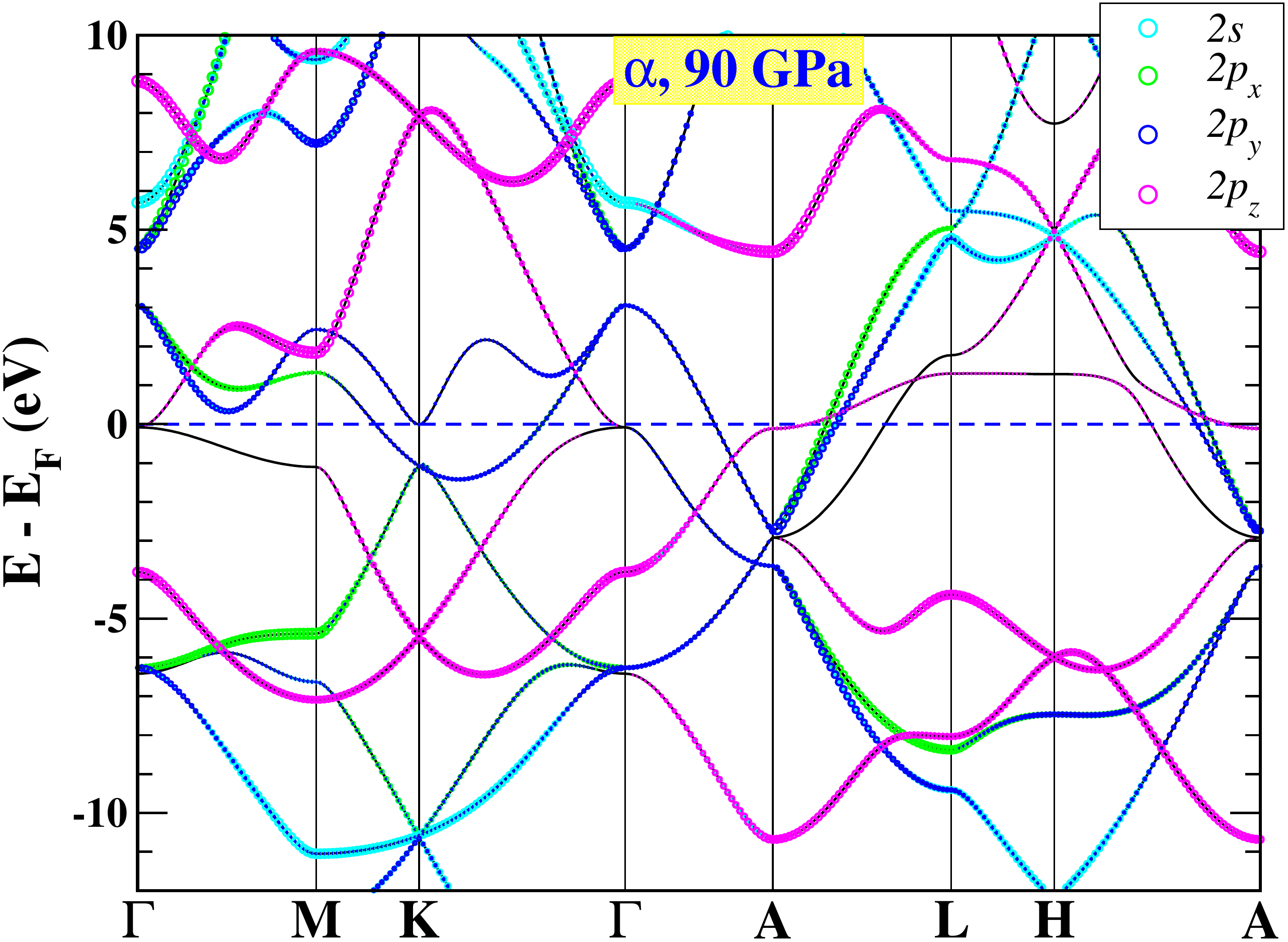}
   \caption{Fatbands representation for $\alpha$-MoB$_2$ at 90 GPa, with top and bottom panels emphasizing Mo and B character, respectively. Mo $4d$ orbitals dominate the Fermi level region. 
   }
   \label{fig:a-fatbands-90}
\end{figure}

\section{Electronic structure}

\begin{figure}[!ht]
  \includegraphics[width=\columnwidth]{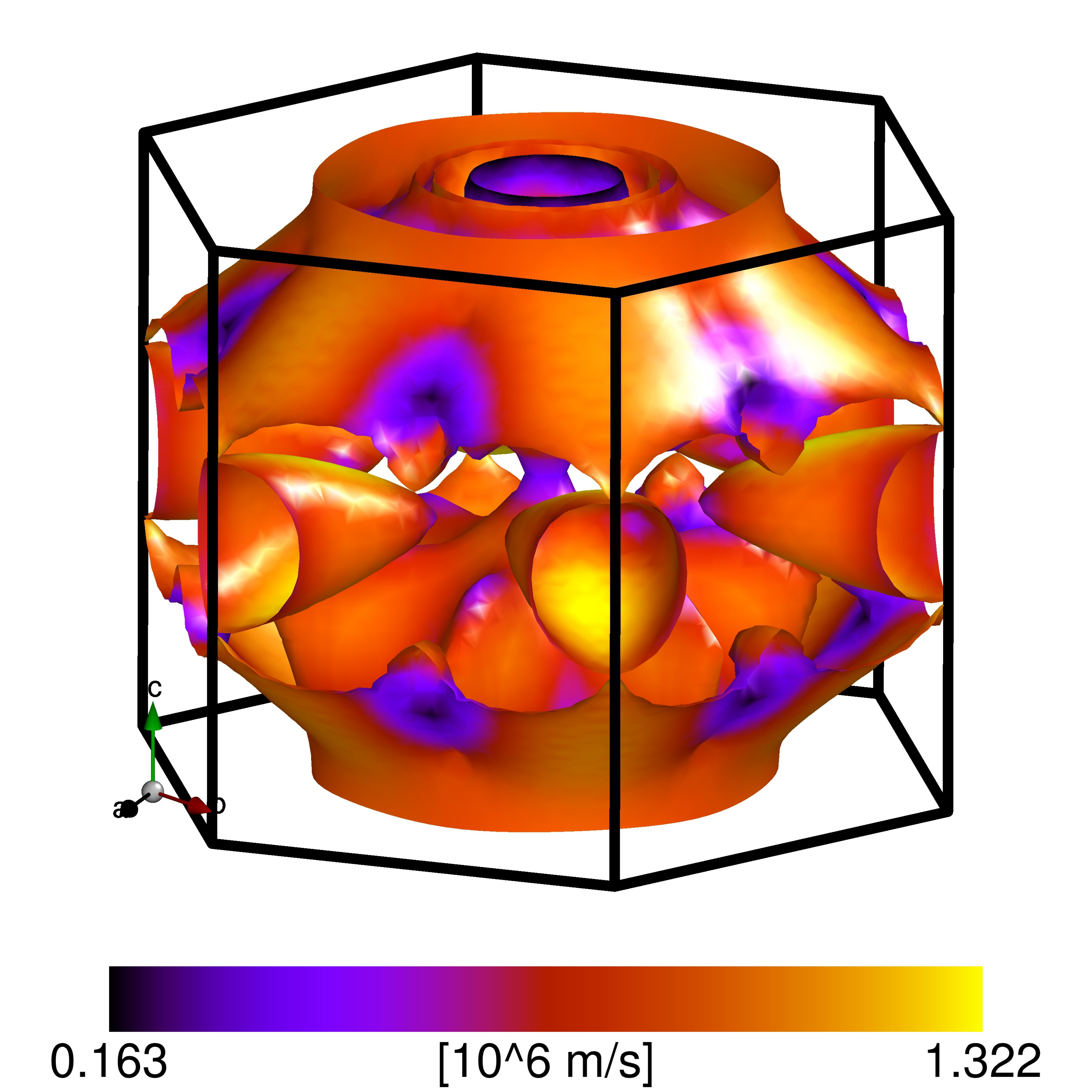}
  \caption{Fermi surfaces of $\alpha$-MoB$_2$ at 90 GPa, showing several large, three dimensional surfaces. Color provides the velocity variation around the Fermi surfaces.}
  \label{fig:a-fs-90}
\end{figure}

\begin{figure}[tbp]
  \includegraphics[width=\columnwidth]{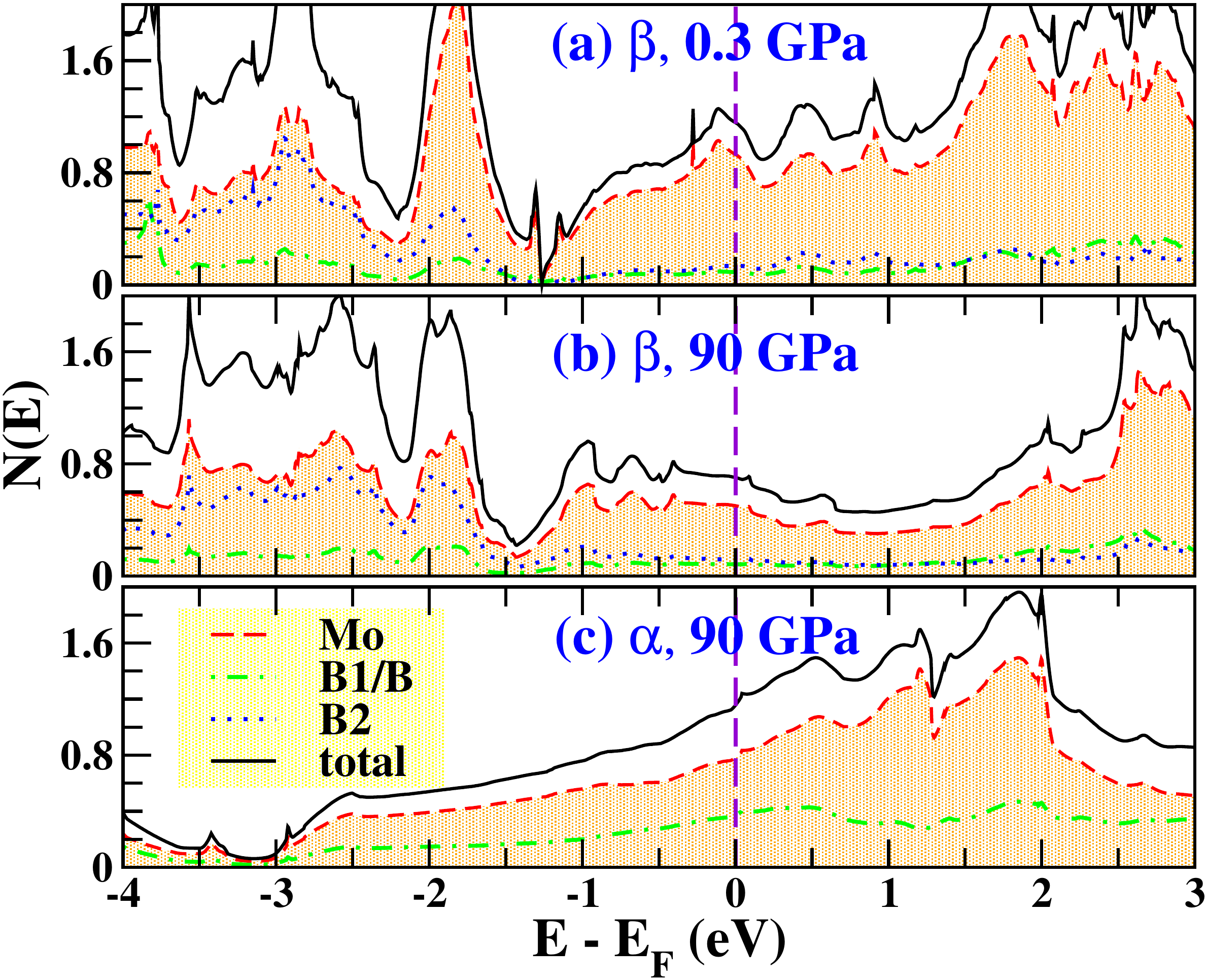}
  \caption{Total and atom-projected electronic densities of states (DOSs) per formula unit for the $\beta$-phase at (a) 0.3 GPa and (b) 90 GPa, and (c) the $\alpha$-phase at 90 GPa.
  The total DOS $N(0)$ at the Fermi energy $E_F$ are (a) 1.16, (b) 0.70, and (c) 1.16, in units of states per eV.
The Mo contributions to $N(0)$ are about 80\% for (a) and 70\% for (b) and (c).
 Here, the experimental structural parameters are used in {\sc fplo}.}
  \label{fig:3pdos}
\end{figure}

We focus on the highest T$_c$ material -- $\alpha$-phase at 90 GPa -- with a few results for the critical pressure $P_c$=60 GPa of the $\beta-\alpha$ structural transition. Several results for the $\beta$-phase are presented in the Appendix. 
The 90 GPa band structure of $\alpha$-MoB$_2$, {\it i.e.} the high $T_c$ phase, is shown in Fig.~\ref{fig:a-fatbands-90} in orbital-projected fatbands fashion.  The bands around E$_F$, and in the range -5 eV to +5 eV, are predominantly Mo $4d$, with the bandwidth strongly modulated by coupling to B states. States threading the Fermi level have strong but not predominant $d_{z^2}$ character; all $4d$ orbitals have some contribution at E$_F$. The partially filled B $s$-$p$ bands are somewhat repelled from the Fermi level region. This behavior is entirely different from MgB$_2$, where Mg donates its electrons to the B bonding bands crossing E$_F$, leaving negligible Mg participation at E$_F$. This strong difference is not surprising, as an open-shell transition metal is nothing like a electropositive divalent $s$-$p$ atom. 

The Fermi surfaces of $\alpha$-MoB$_2$ at 90 GPa, shown in Fig.~\ref{fig:a-fs-90},
are comprised of three large, three dimensional (3D) sheets without apparent nesting features. The r.m.s. Fermi velocity is $\sim$8$\times 10^{7}$ cm/s, with small regions of a low velocity as low as 3$\times 10^7$ cm/s, and highest values up to 13$\times 10^7$ cm/s. This range of velocities, several large FSs, and 3D character indicate that a $k$-averaged, ``single band'' ({\it i.e.} single gap) characterization should apply.   

The projected densities of states (PDOS) for (a) $\beta$ at 0.3 GPa, (b) $\beta$ at 90 GPa, and (c) $\alpha$ at 90 GPa are displayed in Fig.~\ref{fig:3pdos}(a,b,c) respectively. The values of N(0), equal to (a) 1.16, (b) 0.70, and (c) 1.16 states per eV, respectively, are 70-80\% Mo $4d$. The effect of pressure on $\beta$-MoB$_2$ is primarily to broaden the bands, with some band and DOS rearrangement around E$_F$ that lowers N(0) substantially. Recall that this pressure region is where the observed T$_c$ appears and {\it increases substantially}, to 28 K. 
The structural transition 
from $\beta$ to $\alpha$ phase, where the honeycomb layers flatten, changes $N(E)$ considerably around E$_F$, leaving a smooth increase from -2.5 eV to +0.5 eV. Experimentally, T$_c$ continues to increase \cite{MoB2}, but more modestly, above this pressure to 32 K.

\begin{table*}
	\begin{tabular}{|c|cccc|rcc|cr|cr|}
	\hline\hline
	& $N^{\uparrow}(0)$ & ${\cal I}^2$ & $\eta$ 
	          &  $\sqrt{{\cal I}^2}$ 
	              & $\omega_{log}$ & $\omega_2$ 
	                & $\kappa$=$M\omega_2^2$ & ${\cal A}$ & T$_c^{LC}$  & $\lambda$ & $T_c^{AD}$ \\
	& ~(1/eV-spin)~ & ~(eV/\AA)$^2$~ &~(eV/\AA$^2$)~& (eV/\AA) & ~(meV)~ & ~(meV)~ & (meV/\AA$^2$) & (meV)& (K) &  &~(K)~\\
	\hline
	Total & 0.58 &      &      &     & 21.8 & 36.1 &      & 23. & 39.5 & 1.67 & 33.7 \\
	Mo    & 0.38 & 35.1 & 13.6 & 5.9 & 18.2 & 19.7 &  9.9 & 14. & 24.0 & 1.48 & 23.6 \\
	B$_2$ & 0.20 & 21.2 & ~4.6 & 4.6 & 85.2 & 91.5 & 23.1 & 9.  & 15.5 & 0.19 & ~0.0 \\
	\hline\hline	
	\end{tabular}
		\caption{Superconductivity parameters of $\alpha$-MoB$_2$ (AlB$_2$ structure) at 90 GPa. $N^{\uparrow}(0)$ is per formula unit.
	${\cal A}$ is the Leavens-Carbotte area ${\cal A}$ of $\alpha^2F(\omega)$, and T$_c$ is for $\mu^\ast$=0.13. We use the most common isotopic masses (in amu) $M_{\rm B}$=11, $M_{\rm Mo}$=96 (this latter value is the abundance-weighted value of several stable isotopes). Blank spaces indicate there is no such quantity defined for the compound.}
	\label{table:lambda-alpha}
\end{table*}

\newpage
\begin{figure}[!ht]
\includegraphics[width=0.95\columnwidth,height=8cm]{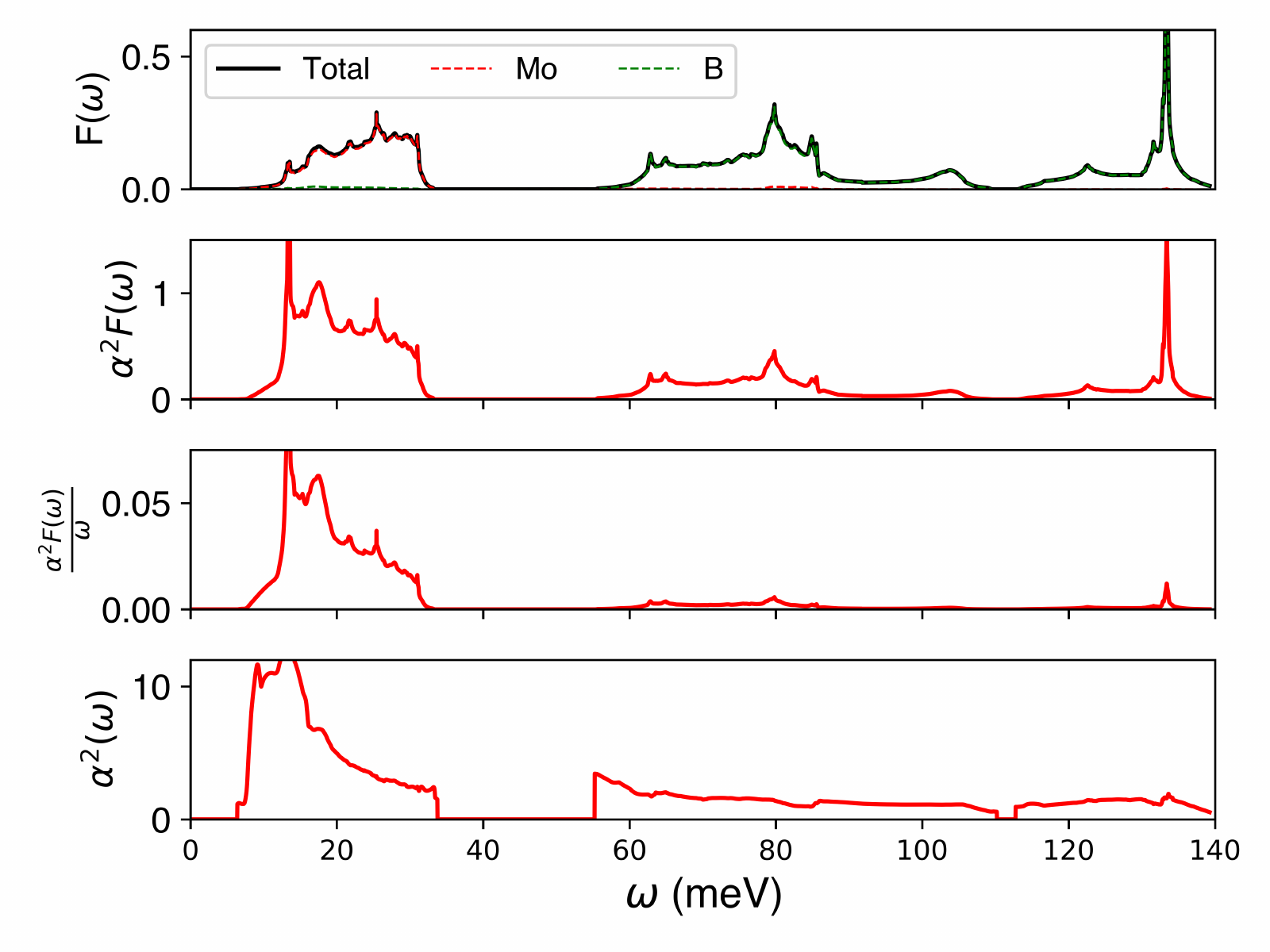}
  \caption{For $\alpha$-MoB$_2$ at 90 GPa: top panel, the atom-projected phonon density of states $F(\omega)$, in meV$^{-1}$; second panel, $\alpha^2F(\omega)$, whose integral gives the area ${\cal A}$ in the Leavens-Carbotte expression for $T_c$; third panel,  $\alpha^2F(\omega)/\omega$ (meV$^{-1}$), whose integral gives $\lambda$; bottom panel, $\alpha^2(\omega)=\alpha^2F(\omega)/F(\omega)$ (in meV). The acoustic mode (Mo) region is dominant except in $F(\omega)$.}
  \label{fig:a-a2F-90}
\end{figure}

\section{Phonon dispersion and electron-phonon coupling}
\subsection{The frequency spectrum}
The atom-projected phonon density of states (phDOS), shown in the top panel of Fig.~\ref{fig:a-a2F-90}, reveals a key aspect of our analysis: the Mo and B modes are disjoint to very good approximation: Mo (acoustic) modes in the 15-35 meV range, and B (optic) modes from 60 meV to 140 meV, with a gap at 115 meV between the lower four and the upper two branches. This atomic separation allows the formalism of Sec. II to apply. Also shown are $\alpha^2F(\omega)$, $\alpha^2F(\omega)/\omega$, and the ratio $\alpha^2F(\omega)/F(\omega)=\alpha^2(\omega)$. 

In $\alpha^2F(\omega)$ the weight is divided 60-40 (Mo, B). This weight is the area ${\cal A}$ emphasized by Leavens and Carbotte \cite{leavens}, as a simple yet reasonably reliable indication of importance leading to their empirical equation
\begin{eqnarray}
T^{LC}_c \approx 0.148 {\cal A}.
\label{eq:LC}
\end{eqnarray}
The area ${\cal A}$ for each atom (Mo, B$_2$) is given, along with the other quantities that are important for superconductivity, in Table \ref{table:lambda-alpha}. The total area ${\cal A}$=23 meV gives T$^{LC}_c$=39 K: 24 K from Mo, 15 K from B. This is a noticeable overestimate given the previous accuracy of this expression, which can be ascribed to the overestimate of the contribution of B. For reference, in five high T$_c$ metal hydrides over three structural classes at high pressure \cite{Quan2019}, the Leavens-Carbotte expression gave impressively good estimates over a range of T$_c$ values of 180-280 K. We now extend this comparison of Mo and B contributions.

\subsection{Importance of Mo versus B$_2$}
Here we provide a few ways of interpreting the relative importance for T$_c$ of the low and high frequency regions of the spectrum.
In $\alpha^2F(\omega)/\omega$, whose integral gives $\lambda$/2, the low frequency Mo weight dominates mightily, within $\lambda_{\rm Mo}/\lambda_{\rm B_2}$=7.5 (see Table~\ref{table:lambda-alpha}), or 15 times larger for Mo than for each B. From the frequency moments given in the table, the net (full spectrum) gives $\omega_{log}$=
21.8 meV, $\omega_2$=
36.1 meV, the ratio of 
1.66 indicating a large ``shape'' aspect to the spectrum and resulting coupling. For each atom separately, the shape ratio is: Mo, 1.08; B, 1.07. This small shape factor for B, and the moments themselves, indicate almost negligible effect of the upper two-thirds of $\alpha^2F(\omega)/\omega$ in the B region (which is hardly visible in Fig.~\ref{fig:a-a2F-90}).

Boron contributes ({\it i.e.} adds to the Mo contribution) a 13\% increase in $\lambda$, a 20\% increase in $\omega_{log}$, and a much larger 83\% increase in $\omega_2$. Recall that $\omega_{log}$ ``sets the scale'' for T$_c$ (in the Allen-Dynes T$_c$ equation), while the increase in $\omega_2$ increases the denominator $\kappa$ in $\lambda$ by its square. Including the B contribution fully does offset this somewhat by decreasing the mass factor in $\kappa$; because it is a compound one cannot be more quantitative, except that the increase in $\lambda$ finally is only 13\%.  On balance, it is clear that B plays a crucial role in the resulting 43\% increase in $T_c$ compared to the value for Mo alone, in spite of its seeming near irrelevance in the $\alpha^2F(\omega)/\omega$ plot. 

MoB$_2$ is thus another example, from the high T$_c$ metal hydrides and several other viewpoints, that so far, raising T$_c$ has been very largely a matter of getting coupling at high frequency, or at least some of the coupling. For the past 50 years, the highest known EP superconductors have had $\lambda$ not much larger than two, whereas T$_c$ has increased by a factor of 12 (23 K to $\sim$280 K). Meanwhile there has been little progress in reaching the ``large $\lambda$'' regime, which is typically considered to be $\lambda > 5$.  The importance of increasing $\lambda$ has been seriously overvalued in the quest to raise T$_c$ \cite{Pickett2021}.

\subsection{Frequency dependence of the coupling}
The spectra in Fig.~\ref{fig:a-a2F-90} are informative. The phDOS $F(\omega)$ is about as might be expected given the nearly factor of nine between the masses. The acoustic modes, up to 35 meV, are essentially all Mo character, the optic (B) modes run from 60 meV to 140 meV. They include three and six branches, respectively. The B modes are split into bond-stretching modes above the small gap at 110 meV, with the bond-bending modes lying below. $\alpha^2F(\omega)$ shows a relative strength in the Mo region and weakness for B. The plot of $\alpha^2F(\omega)/\omega$, which integrates to $\lambda/2$, makes vivid the dominance of Mo for $\lambda$, from which this dominance might be thought to extend to T$_c$. As discussed above, the understanding is not really so simple: coupling to B provides almost one-third of T$_c$ when viewed as being on top of the Mo contribution. The B contribution is difficult to see but finally integrates to something useful.

The fourth (lowest) panel provides $\alpha^2(\omega)$, how the strength of coupling -- including mass and frequency dependence in the matrix elements but independent of the phDOS -- is distributed. It is strongest in the (Mo) acoustic region, being strongest in the lower third of the frequency range (the data in these curves below 6 meV is missing due to the numerical cost of handling the low phase space region near the $\Gamma$ point, which is unimportant for T$_c$ anyway). In the B part of the spectrum, $\alpha^2(\omega)$ is largest in the lower region, slowly decreasing then leveling off. Notable is that $\alpha^2(\omega)$ is no different for stretching versus bending modes; in MgB$_2$ the bond stretches are all-important. The average of $\alpha^2(\omega)$ in each region is related to the electron-ion matrix elements discussed next, though not in a simple manner (note the difference in units).

\subsection{Electron-ion matrix elements}
EP matrix elements, whose understanding is still at a primitive stage, consist of an electron-ion scattering strength ${\cal I}$ (its rms value after averaging over the FS) divided by a phonon factor $(M\omega^2)^{1/2}$ (for elements, or for disjoint atomic modes such as in MoB$_2$). At the core is the change in self-consistent electronic potential upon small displacement of an atom, Eq.~(\ref{eq:matrixelement}). This change requires the static electronic response, which is available in electronic structure codes from self-consistent linear response methods, but is not yet extracted in any enlightening way. What we find here is that for MoB$_2$, 
\bea
{\rm Mo}: {\cal I}=5.9~ {\rm eV/\AA};~~~~{\rm B}: {\cal I}=4.6~ {\rm eV/\AA}.
\label{eq:I}
\eea
These are averages over all modes of each atom, obtained because $\lambda$, $M\omega_2^2$, and $N^{\uparrow}(0)$ are known for each atom. Further discussion is provided in the next section.

These can be compared with high pressure metal superhydrides, with T$_c$ in the 175-275 K range. In the hydrides \cite{Quan2019} it was found that systematics could be obtained if the metal atom contribution was neglected: it raises $\lambda$ but decreases frequency moments, with little -- and sometimes negative -- effect on T$_c$.  Values of ${\cal I}_{\rm H}$ ranged from 4.5 eV/\AA~to 10.5 eV/\AA, with values correlating positively (but only roughly) with T$_c$. There was also a relatively small but positive correlation with increasing pressure, all features yet to be understood because ${\cal I}$ is a derived quantity in this analysis, not separately calculated and analyzed although the capability is essentially in place in existing codes. A simple picture would suggest that higher $N^{\uparrow}(0)$ would lead to more static screening hence a smaller value of ${\cal I}$, but study of such relationships remains to be done.

\subsection{Isotope shift of T$_c$}
Using the formalism of Sec. II.C and our calculated data, we have calculated the individual isotope shift coefficients $\alpha_{iso}$ for Mo and B isotope replacements individually. In BCS theory where there is a single elemental mass, no Coulomb $\mu^*$, and no strong coupling or shape factors, this shift is $\frac{1}{2}$, corresponding to the dependence of the phonon energy scale on the square root of the mass. The calculations were done for $\mu^*$=0.13 as in the rest of this study. This value is just a choice as in the middle of the common 0.10-0.15 range. We repeated calculations of this section for $\mu^*$=0.10, and T$_c$  by more than 10\%. Thus the choice of $\mu^*$ must be considered when making absolute comparisons with (say) experiment, whereas trends will not be much affected.

The values obtained are 
\bea
\alpha_{iso}^{Mo}=0.435;~~~~ \alpha_{iso}^B=0.071. 
\label{eq:isotope}
\eea
(We note that these values are changed by -0.003 and +0.003 respectively for $\mu^*$=0.10.) These values follow the general expectation: T$_c$ is mostly due to the Mo contribution to $\alpha^2(\omega)$, much less to the B contribution although it does boost T$_c$. The changes, for 1 amu change in mass, in the atomic values of $\omega_{log}$ and $\omega_2$ respectively are about 0.5\% for Mo (0.1 meV), and 5\% (4.5 meV) for B, while $\lambda$ and the individual atomic contributions to $\lambda$ of course do not change. 

A point of interest pertains to the strong coupling $f_1$ and shape $f_2$ factors in the T$_c$ equation \cite{alldyn}; recall that $\lambda$=1.67 is strong coupling, and the shape of $\alpha^2F(\omega)$ is bifurcated. These values are $f_1$=1.09, $f_2$=1.056, together giving a 15\% larger value of T$_c$ than from the `modified McMillan equation' \cite{alldyn} where these factors are not present.

\section{Why {\rm M\lowercase{o}B$_2$} at 90 GP\lowercase{a} is a HTS}
\subsection{Background: historical studies}
Except for the compressed metal hydrides, electron-phonon superconductors rarely display T$_c$ above 23 K (even under pressure), with MgB$_2$ being the exception. (The mechanisms for metal fullerides and (Ba,K)BiO$_3$ remain unclear, as they may involve Coulomb contributions.) All of the atom-specific factors in $\lambda$ are available, and the phonon frequency spectrum and coupling to obtain $\alpha^2F(\omega)$ have been calculated. This enables one to begin to make comparisons with other good superconductors, although such full detailed data is often not available from the publications. We consider Mo in the context of transition metals in the middle of the periodic table, which have dominated the high T$_c$ conventional superconductors.

The conundrum of raising T$_c$ has a substantial, and informative, literature. One has a coupling strength $\lambda$ that is to be increased, and a phonon energy scale -- we'll use $\omega_2$ here -- to be increased. However, $\lambda\propto 1/\omega_2^2$ so increasing the frequency decreases coupling strength, as  the square. Increasing $\lambda$ enhances the renormalization downward of $\omega_2$ due to EP coupling, thus decreasing $M\omega_2^2$ and with many examples of the lattice being driven unstable as T$_c$ increases. This renormalization involves electronic screening, with few general and useful guidelines having emerged. In the expression for $\lambda$ in Eq.~(\ref{eq:lambda}), $N(0)$ and $M\omega_2^2$ have received a great deal of attention, with the scattering strength ${\cal I}^2$ remaining mostly a mystery.
Also, the strong coupling limit \cite{alldyn} T$_c \propto \sqrt{N(0){\cal I}^2/M}$ focuses attention on ${\cal I}$, and Allen and Dynes noted that high T$_c$ superconductors at that time achieved it by means of large values of $\eta$.

Essentially all studies of ${\cal I}^2$ have applied the Gaspari-Gy\"orffy (GG) model \cite{GG} of a rigidly displaced atomic potential treated with scattering theory (rigid muffin-tin approximation). The immediate result of the formalism is a scattering amplitude for each atomic partial wave (angular momentum $\ell$ value 0, 1, 2, 3), weighted by corresponding partial $N(0)$ values to give $\eta=N^{\uparrow}(0){\cal I}^2$ and thereby ${\cal I}$. Hopfield suggested $\eta$ as an atomic property \cite{Hopfield}, but it is ${\cal I}$ that is more nearly so since $N(0)$ can vary so strongly depending on band details. Phillips suggested \cite{Phillips1972} that in binary compounds with widely differing masses, the total coupling can be imagined as contributions from the separate atoms, as described in Sec. II. 

The separation outlined in Sec. II was first applied by Klein and Papaconstantopoulos \cite{BMK-DAP-1974} to V, Nb, and (Nb,Ta,Hf)C, all superconducting except the latter. Nb, neighboring Mo in the periodic table, had (all in eV/\AA) elemental ${\cal I}$=2.9 where T$_c$=9.2 K, and 4.5 in the carbide (T$_c$=11.1 K), a considerable difference especially when squared. The C values, by the way, were very similar to Nb and Ta in the carbides. A following study of Nb and V materials (elements, carbides and nitrides, A15 compounds) \cite{BMK1979} concluded that ${\cal I}$ is reasonably considered as an atomic property, equal to 3-3.7 eV/\AA~ throughout these classes, roughly consistent with the quotes above. In the same classes, the value for V materials was 2.4-2.7 eV/\AA. Adding to these materials an enlarged set of A15 compounds, it was found that, at least at the GG level, for Nb ${\cal I}$=3.0-3.8 eV/\AA, and for V ${\cal I}$=2.3-2.6 eV/\AA, for wide variations in T$_c$.  Thus ${\cal I}$ seems very much to be an atomic property, at ambient pressure \cite{Pickett1982}.

\subsection{Molybdenum}
Now we address Mo. Rocksalt MoN, predicted to be a 30 K superconductor, was calculated from GG theory to have ${\cal I}$=3.5 eV/\AA~ for Mo. It was its very large $N(0)$ that resulted in large coupling and high T$_c$; later it was found calculationally to be elastically unstable \cite{Mehl}, also surely due to its large $N(0)$. Our analysis for MoB$_2$ provides new data about the importance of atomic scattering. Table II provides the value ${\cal I}$=5.9 eV/\AA, significantly larger than the values for Nb, moreover its square is {\it nearly three times larger} than for Mo in MoN. Our observation just above that ${\cal I}$ may be a transferable atomic property does not hold well for Mo when pressure is applied. Perhaps pressure gives important enhancement of ${\cal I}$. Support for this interpretation has been given by Papaconstantopoulos {\it et al.} in their study of H$_3$S \cite{Papa2015}, where for pressures from 50-300 GPa the value of ${\cal I}$ for hydrogen increased from 3 eV/\AA~ to nearly 5 eV/\AA.

The value for B of 4.6 eV/\AA~in MoB$_2$ is similarly impressive, though other factors finally result in weak B coupling, as measured by $\lambda$. This is another indication that the integrand $\alpha^2F(\omega)/\omega$ strongly penalizes coupling at high frequency: $\lambda$ has been overly emphasized in the quest for higher T$_c$ for several decades. This was the crux of the Leavens-Carbotte picture. The relative importance of Mo and B is consistent with observations from earlier in this paper: MoB$_2$ is well characterized as a Mo superconductor enhanced, surprisingly strongly, by some modest coupling to B high frequency modes. How Mo obtains the large value of ${\cal I}$, viz. what is operating in the pressure response of the displacement of potential, remains the important question for further study. This seems, given that $N(0)$ is not large, to be the attribute responsibility for 32 K superconductivity in MoB$_2$.


\section{Summary}
MoB$_2$ is one of several transition metal diborides that have been studied, hoping without success to find T$_c$ approaching or possibly exceeding that of MgB$_2$. In fact, MoB$_2$ at ambient pressure is not superconducting. High pressure produces an onset and impressive increase of T$_c$ in the $\beta$-phase, after which a transformation to the higher symmetry $\alpha$-phase (AlB$_2$ structure type) occurs and T$_c$ continues to rise, but more slowly. Like most other TM diborides with this AlB$_2$ structure, MoB$_2$ is a three dimensional, multiple Fermi surface metal with very strong metal $3d$ character. Also similarly, states at and around the Fermi level are primarily Mo $4d$ in character. The theoretical value of T$_c$ at 90 GPa is in excellent agreement with the experimental data.

A main,  and immediate, conclusion here is that MoB$_2$ is not only very different from MgB$_2$ as anticipated, it is also different from the metal hydrides with T$_c$ approaching room temperature under pressure, with analysis facilitated \cite{Quan2019} by the large mass difference of the atoms, as occurs in MoB$_2$. Like the hydrides, MoB$_2$ has disjoint metal and light atom phonon spectra that enable simplification of analysis of the origins of electron-phonon coupling and superconductivity. However, in the hydrides it is best to neglect the mutually compensating metal atom contributions \cite{Quan2019}, after which understanding arises from the metallic hydrogen properties alone \cite{Quan2019}.  In MoB$_2$, the two atom types cooperate to produce the observed 32 K superconductivity. We have found that even weak coupling from the B atoms contributes substantially, by increasing frequency moments with minor coupling strength that raises T$_c$ by 43\%. Individual atomic contributions to $\lambda$ can be misleading when wanting to understand contributions to T$_c$. In hydrides, the metal contribution of $\sim$15\% is useless; in MoB$_2$, the B contribution of a similar fraction is important, because it arises from high frequency.

These results -- the experimental data and theoretical analysis -- encourage future studies of TM diborides at higher pressure. Some of them should display a larger value of $N(0)$, which is a positive contributor until it promotes electronic (magnetic, structural) instability and thus return to a less favorable material. Future work should include more attention to the electron scattering amplitude ${\cal I}$, which our analysis as been able to unravel from other calculational data. Is Mo the best element for providing a large value of ${\cal I}$?

The same separation of the phonon spectrum into atomic contributions allows simple calculations of the isotope shift of T$_c$ for each atom, without full recalculation of phonon frequencies and matrix elements. It is found, as likely expected, that the Mo isotope shift is dominant and close to the BCS value, while the B shift is minor. This separation corresponds more closely to the atomic contributions to $\lambda$ than to the impact of the coupling on T$_c$. Given that B boosts T$_c$ so effectively, it is somewhat of a puzzle why its isotope shift is so small.
 
The analysis provided by Sec. II has revealed the source of the high T$_c$ of MoB$_2$. The Fermi level density of states is ordinary, the phonon spectra as summarized in the atomic values of $M\omega_2^2$ are not unusual. What is distinctive is the electron-displaced atom scattering factor ${\cal I}^2$, which is 2-3 times larger than in zero pressure compounds based on comparable transition metal atoms (V, Nb, Mo) in other compounds. The study of ${\cal I}$ is the primary, perhaps the only, factor yet to be understood for conventional superconductors \cite{Pickett2021}.

The calculated behavior of the $\beta$-phase T$_c$(P) is an item for future study. The observed increase in T$_c$, from zero to 28 K for pressures of 0.3 GPa to 60 GPa, raises interesting questions, more so because we calculate a strongly decreasing $N(0)$ in this same pressure range. It seems that this increase in T$_c$ must be due partially to an increase in frequency moments, but may require a strong increase in matrix elements as well. $N(0)$ is, after all, one of the important factors in $\lambda$  and it is unclear how a strong decrease in $N(0)$ can be compensated by other factors in $\lambda$. Evidently confirmation of the observed behavior by another experimental group, preferably with somewhat different probes, is highly desirable. 

\section{Acknowledgment} We acknowledge Yanpeng Qi for providing and clarifying unpublished structure information. D. A. Papaconstantopoulos provided useful comments on the manuscript.  Y.Q. thanks Stony Brook Research Computing and Cyberinfrastructure, and the Institute for Advanced Computational Science at Stony Brook University for access to the innovative high-performance Ookami computing system, which was made possible by National Science Foundation grant 1927880. Y.Q. and W.E.P. have also used the computational resources on the Pittsburgh Supercomputing Center machine bridges2, supported by NSF TG-DMR 180112. 
K.W.L. was supported by National Research Foundation of Korea Grant No. NRF2019R1A2C1009588.
W.E.P acknowledges support from U.S. National Science Foundation Grant DMR 1607139.

\newpage
\section{appendix}
Because superconductivity in MoB$_2$ is a new research area, and especially that the rise in T$_c$ in the $\beta$ phase will attract more study, we provide more results on the basic electronic structure: band structure, density of states, and Fermi surfaces.  

\begin{figure}[b]
  \includegraphics[width=\columnwidth]{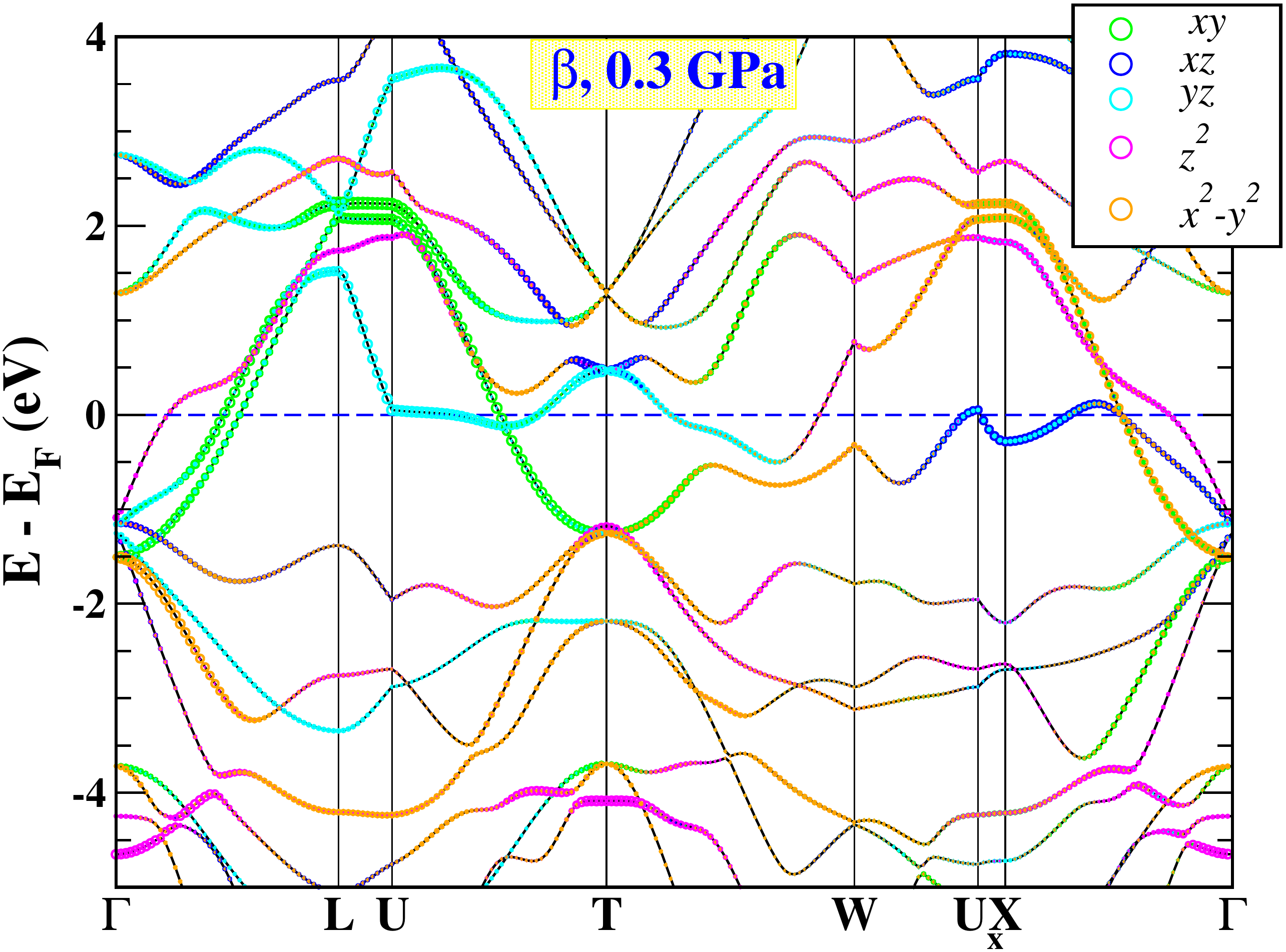}
  \includegraphics[width=\columnwidth]{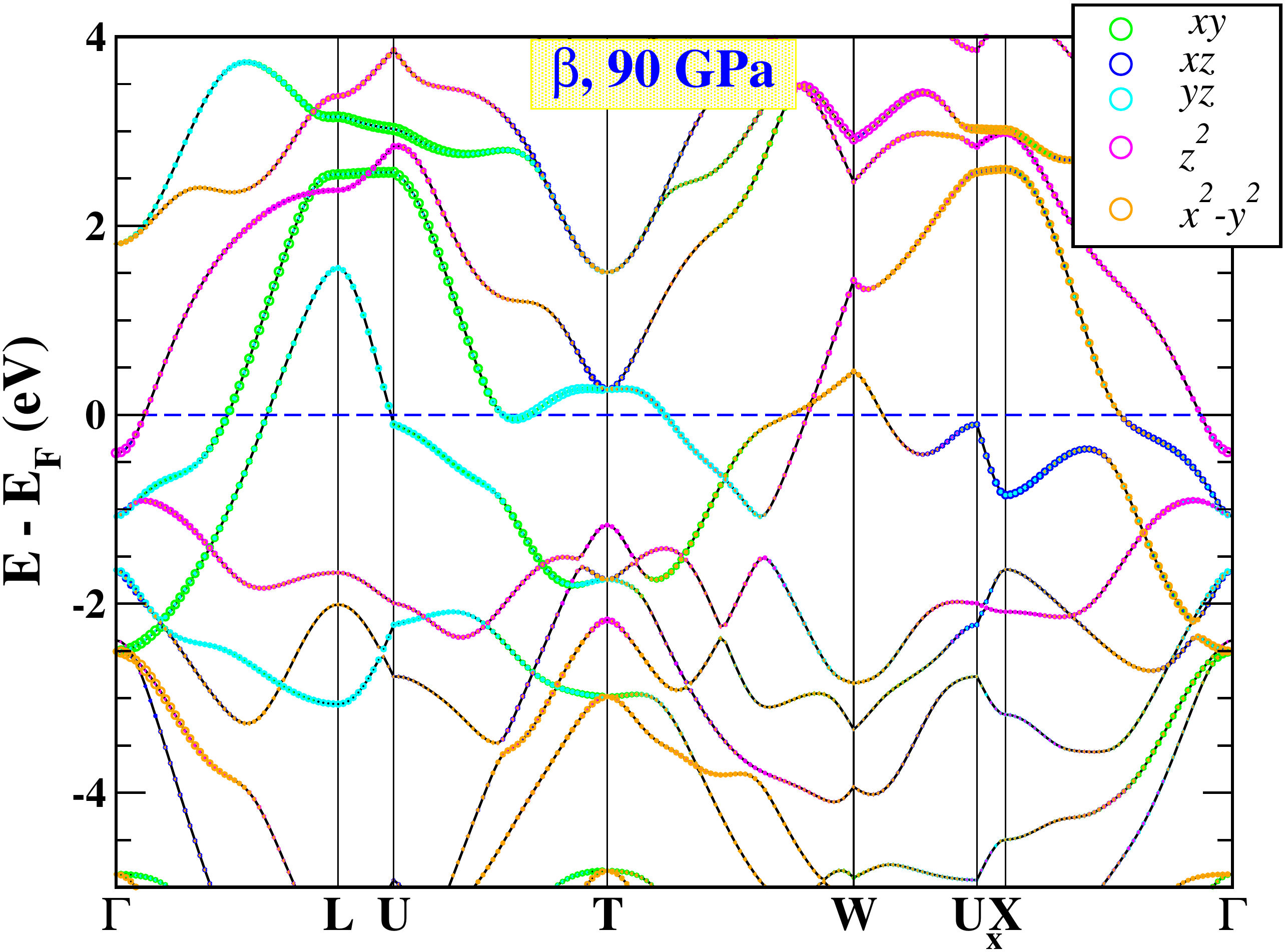}
  \caption{Fatband representation of the Mo $4d$ character in the band structure of $\beta$-MoB$_2$: (top) 0.3 GPa; (bottom) 90 GPa. Here, the experimental structural parameters of {\sc fplo} are used. 
  }
  \label{fig:b-bands-60}
\end{figure}

In Fig.~\ref{fig:b-bands-60} the band structures of the rhombohedral $\beta$-phase are compared at low and at 90 GPa pressures. While there are many detailed changes, in both cases there are multiple Fermi surfaces, including large ones. The increase in bandwidth with volume reduction is not so obvious in the figure, but the states at $\Gamma$ reveal that the $\vec k$=0 eigenvalues spread in energy, and several have disappeared below the panel of 90 GPa bands. The average of the density of states $N(E)$ will decrease accordingly. 

\begin{figure}[!ht]
  \includegraphics[width=\columnwidth]{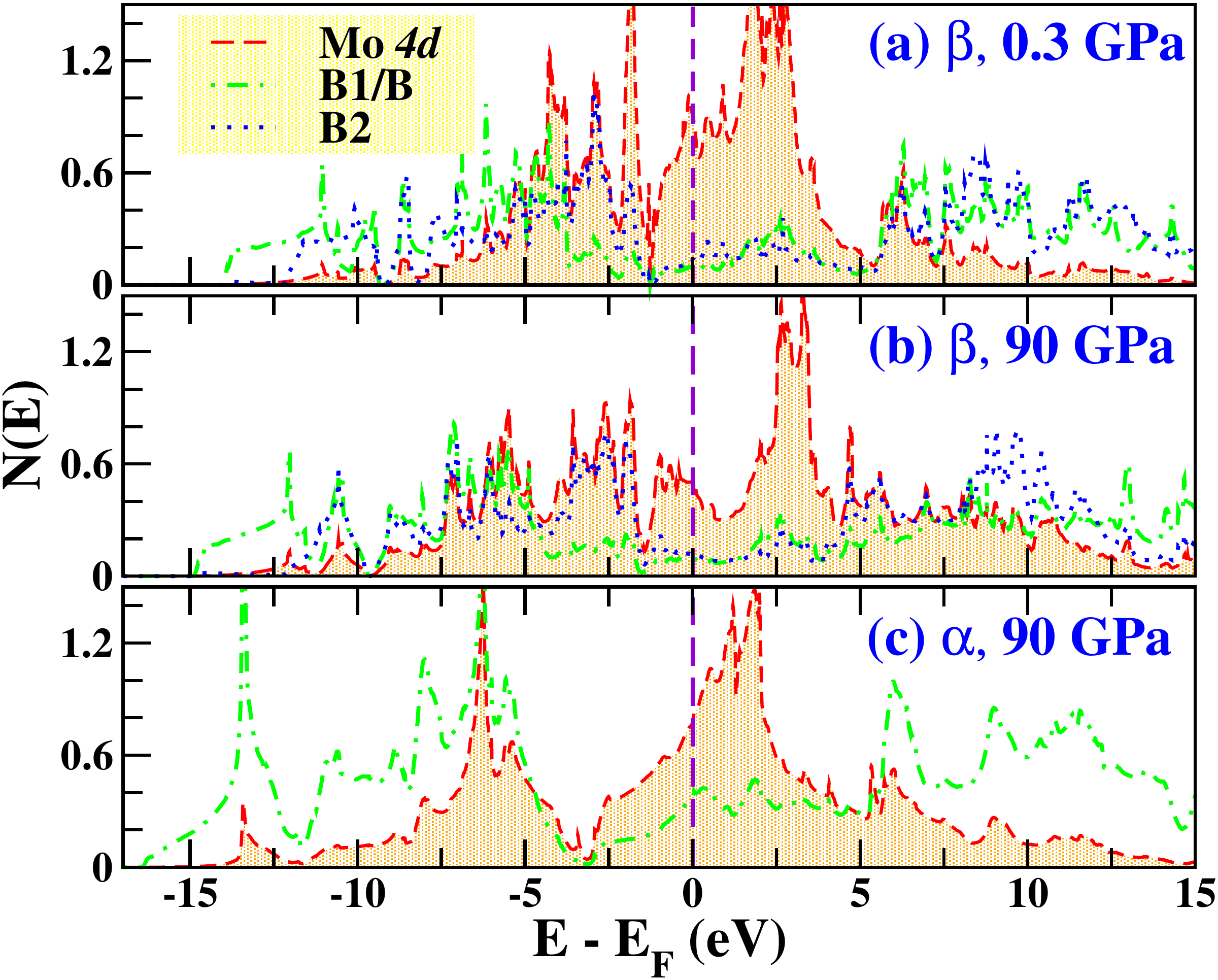}
  \caption{Mo $4d$-orbital and B-atom projected electronic densities of states per eV and per formula unit, in the full valence-conduction range --17 eV to 15 eV, for the $\beta$-phase at (a) 0.3 GPa and (b) 90 GPa, and (c) the $\alpha$-phase at 90 GPa.
  Here, the experimental structural parameters are used in {\sc fplo}.
Pressure in the $\beta$ phase produces the expected band broadening, but also rearrangement in the Mo DOS that {\it reduces} $N(0)$ considerably. Transformation to the $\alpha$ phase restores the Mo DOS to more similar to the low pressure $\beta$ value. }
  \label{fig:opdos}
\end{figure}

The corresponding DOSs in Fig.~\ref{fig:opdos} provide the basic picture. These results indicate, first, that there are rather modest changes in the $\beta$-phase given these are at low and 90 GPa pressure. The change in $N(0)$ is however substantial, being about half at high pressure where T$_c$ is observed to have increased incredibly. Since nominally $\lambda$ is proportional to $N(0)$, this change is unexpected and unexplained (our focus in the main text has been on the $\alpha$-phase). Large changes in the frequency spectrum and coupling are needed to clarify the origin of the emergent superconductivity. 

The structural transition occurs primarily at 60 GPa and certainly is complete at 90 GPa.  The difference between $\beta$ and $\alpha$ at 90 GPa, also shown in Fig. \ref{fig:opdos}, are obvious, with changes in position of large spectral density and a large increase in $N(0)$. Since no discontinuity in T$_c$(P) is apparent, this area is another one of interest for further study.

\begin{figure}[ht]
  \includegraphics[width=\columnwidth]{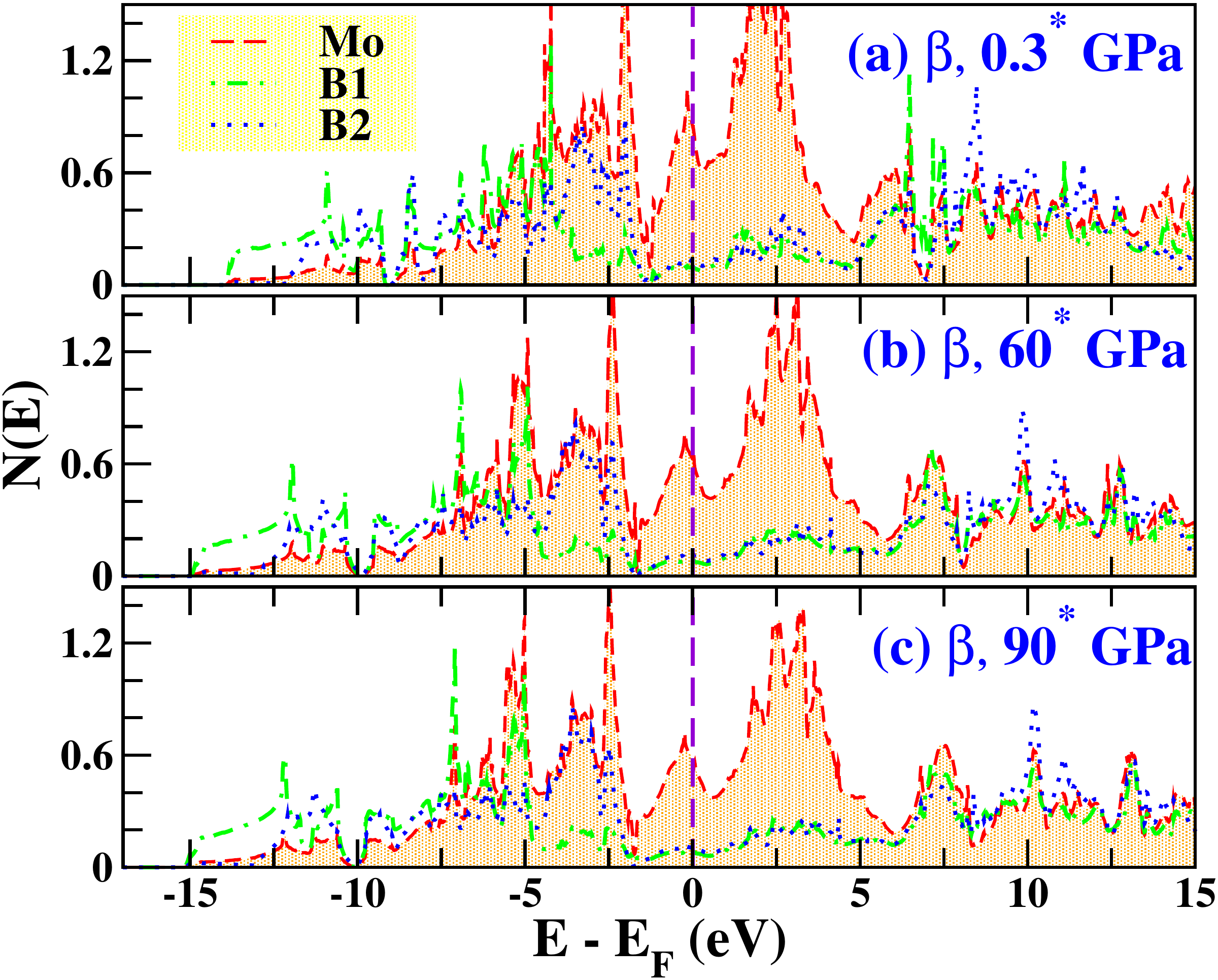}
  \caption{Atom-projected electronic densities of states per eV and per formula unit, in the full valence-conduction range --17 eV to 15 eV, for the $\beta$-phase at 0.3, 60, and 90 GPa. Here, we used the relaxed internal parameters with the experimental lattice constants given in Table \ref{str}.
  Pressure leads to broadening and reducing $N(0)$, but changes between 60 GPa and 90 GPa are negligible.
 }
  \label{fig:Rxpdos}
\end{figure}

Figure~\ref{fig:Rxpdos} shows similar curves, but 
we focus on the two representative pressures, the critical pressure $P_c$=60 GPa of the $\beta-\alpha$ structural transition, and $P$=90 GPa, which is representative of the high T$_c$ $\alpha$-phase. structures in the $\beta$-phase, and also comparing the DOS in the $\beta$ phase at 60 GPa and 90 GPa. This figure illustrates, as mentioned in the main text, that within a given structure, the DOS varies regularly with pressure for MoB$_2$. 
The change at the phase transition provides a different picture of the Mo $4d$ bands (which of course are hybridized with B $2p$ states). A near-zero in $N(E)$ at -3 eV suggests a partition between strongly bonding states below, and less bonding states in the region of E$_F$. The B $2p$ DOS reflects a similar behavior. This character is likely connected with the buckling of the B honeycomb layers, with their changes in Mo-B and B-B bond lengths provided in the main text.    
For these relaxed structures, the variation of $N(E)$ is regular; interestingly, E$_F$ remains at a sharp peak $N(E)$ while the peak, and $N(0)$, lose amplitude with increasing pressure.

\begin{figure}[!ht]
  \includegraphics[width=\columnwidth]{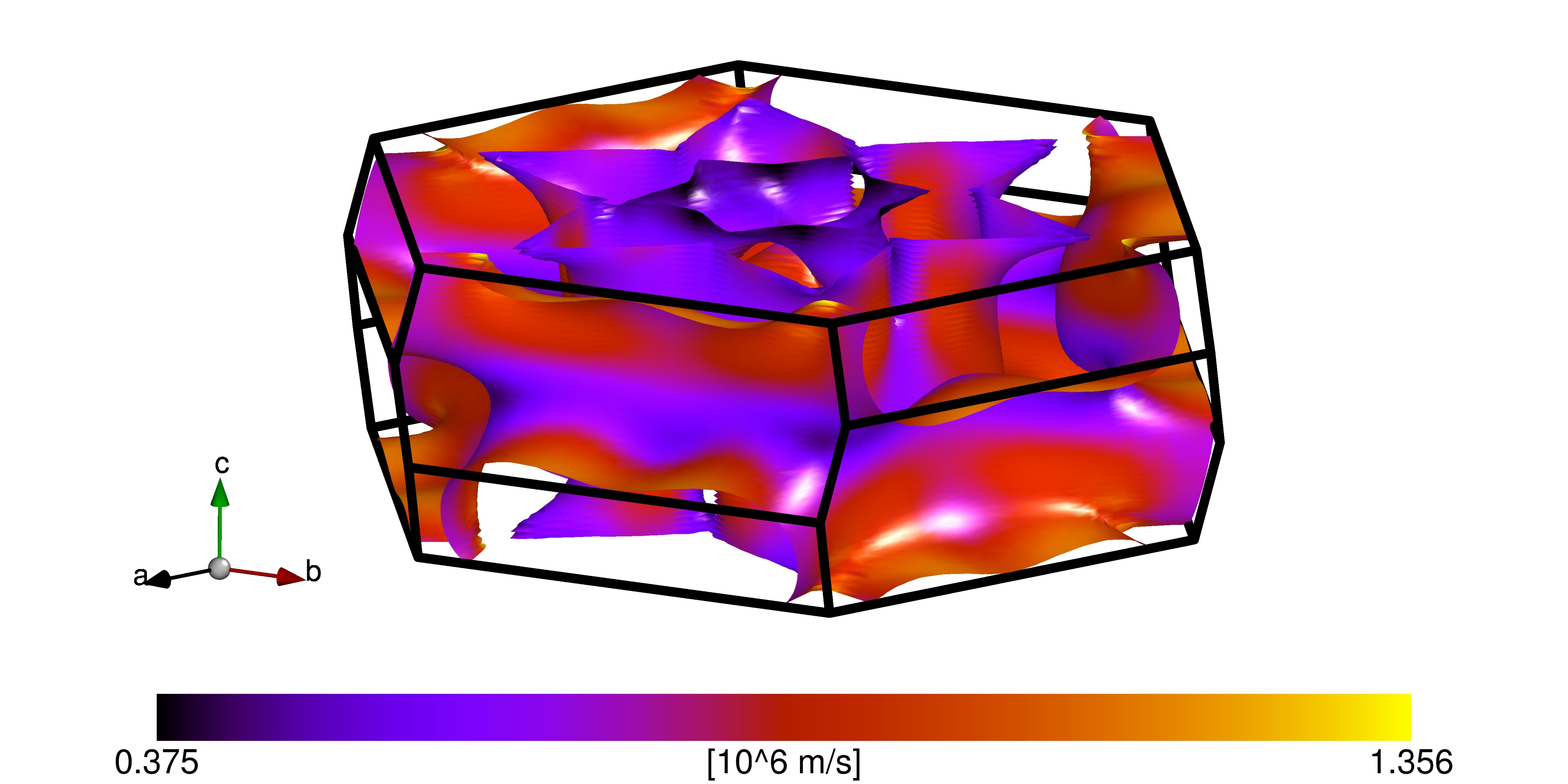}
  \includegraphics[width=\columnwidth]{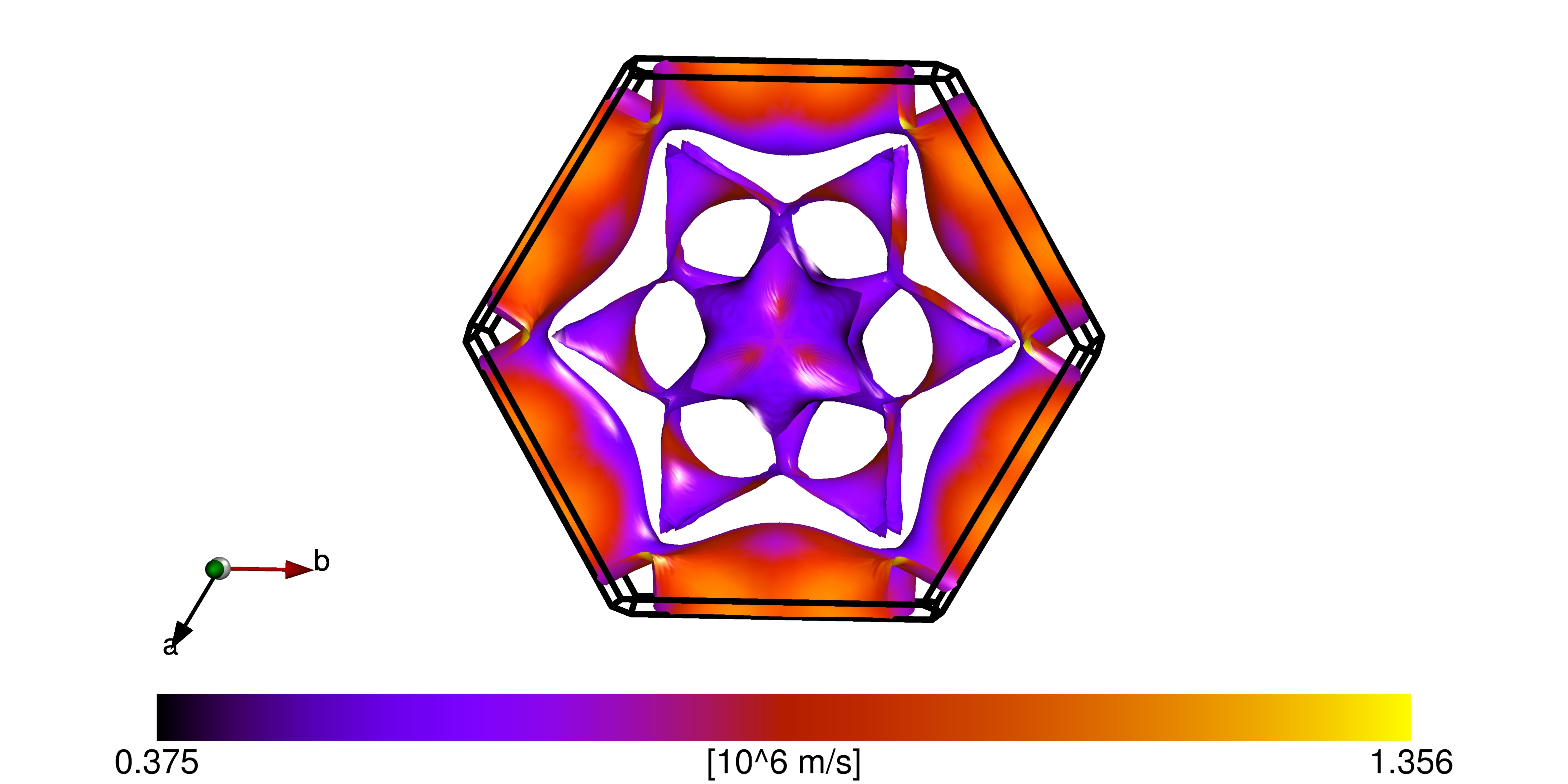}
  \includegraphics[width=\columnwidth]{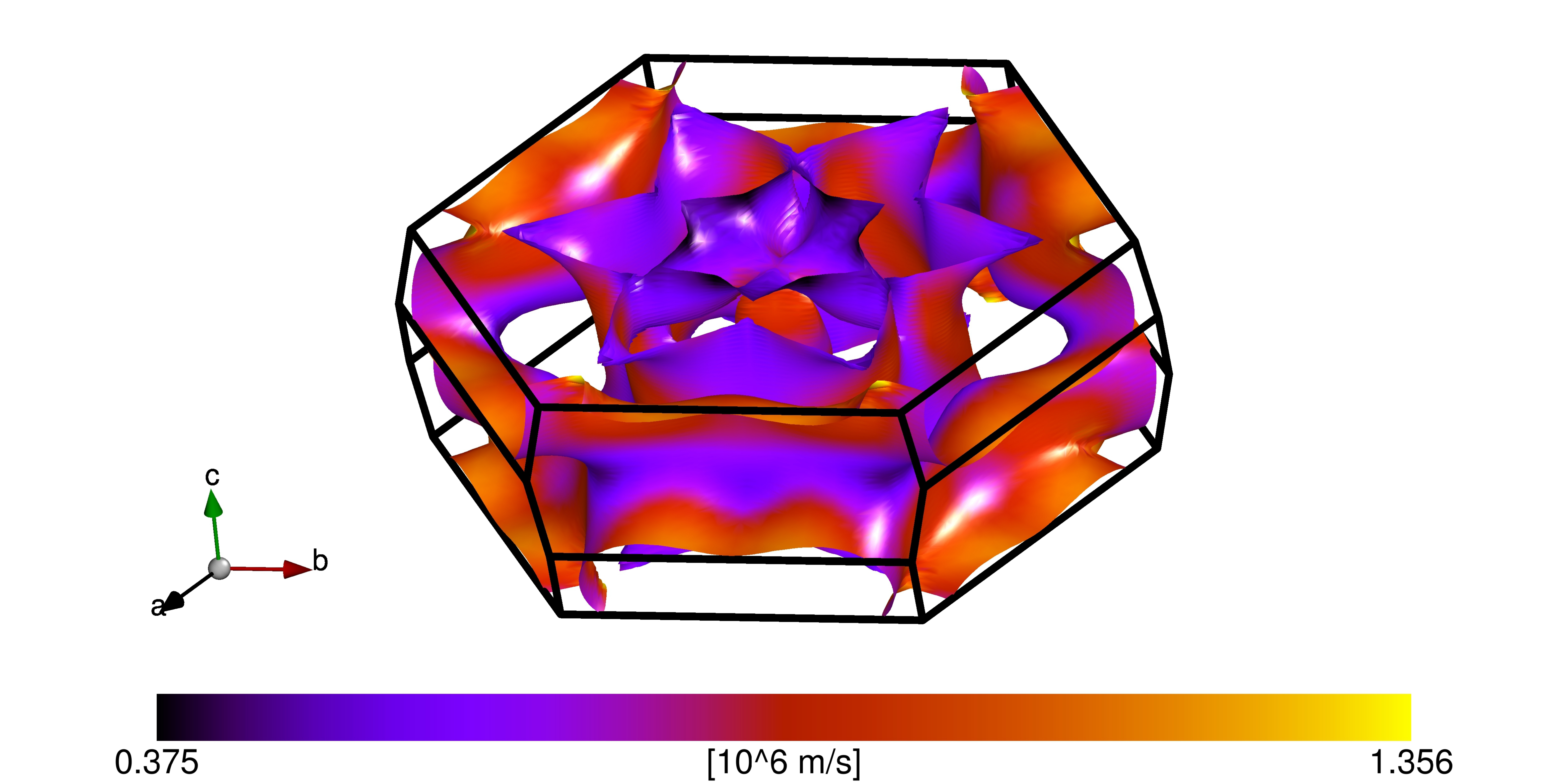}
  \caption{Three views of the Fermi surfaces of $\beta$-MoB$_2$ at 90 GPa with the experimental structural parameters. Color provides the velocity variation around the Fermi surfaces.}
  \label{fig:b-3fs-60}
\end{figure}

Lastly, Fig.~\ref{fig:b-3fs-60} provides three angles of view of the multisheeted $\beta$-phase Fermi surfaces at 90 GPa.  Some two-dimensional character may be arising; the FS has more sheets crossing the top face than the side faces. The top view (center panel) confirms this, as there are ``columns'' of 2D slices with no FS, while the top panel -- a near side view -- shows nothing like this.  The third view provides yet another indication of the degree of two dimensionality. This character, which is not evident from the band structures, is distinct from that of the $\alpha$ phase, discussed in the main text.


\begin{thebibliography}{10}
\bibitem{leavenscomment} The Leavens-Carbotte expressions neglects the $\mu^*$ effect, so consideration of this expression should take that into account.
\end{thebibliography}
\end{document}